\documentclass[11pt, twoside, a4paper, twocolumn]{article}

\usepackage{Build/packages}
\usepackage{Build/commands}
\usepackage{Build/layout}
\usepackage{cite} % PACKAGE TO ORGANIZE CITATIONS
\title{Velocities of mesenchymal cells may\linebreak be ill-defined}

\author{Guilherme S. Y. Giardini$^{a,b}$, Gilberto L. Thomas$^{b}$, Carlo R. da Cunha$^{a}$, Rita M. C. de Almeida$^{b,c,d}$}

\theInstitutions{

$^a$ School of Informatics, Computing, and Cyber Systems, Northern Arizona University, Flagstaff, AZ

$^b$ Instituto de Física, Universidade Federal do Rio Grande do Sul, Porto Alegre, RS, Brazil

$^c$ Instituto Nacional de Ciência e Tecnologia, Sistemas Complexos, Universidade Federal do Rio Grande do Sul, Porto Alegre, RS, Brazil

$^d$ Programa de Pós Graduação em Bioinformática, Universidade Federal do Rio Grande do Norte, Natal, RN, Brazil
}

\begin{document}
\maketitle
 
%Sections
\section{Introduction}
\label{chap:Introduction}

Cell migration is a central mechanism in biological processes including tissue healing \cite{Basan2013,Cui2020}, the immune defense \cite{Qi2018}, embryo development \cite{Alt2017}, and tumor evolution \cite{Wolf2007,Friedl2011,Kabla2012,Szabo2013,Nousi2021}. 

During migration, cells polarize as a result of the creation of gradients of protein concentration along their rear-front axis \cite{Stossel1990,Hall2000,Maree2006,Hennig2020}. These gradients are driven by active molecular transport facilitated by micro-tubule and actin fiber organization\cite{Bugyi2020}. At the leading edge of the cell, actin monomers form a cytoskeletal network that extends against the cell membrane, creating a lamellipodium, which propels the entire cell forward through adhesion to the substrate followed by cytoskeletal contraction \cite{Mogilner2009}. With the assistance of active transport of network components, the cell achieves stable and uniform movement \cite{Shao2012,Mogilner2020}.

Although the dynamics of cell migration is typically uniform along the polarization direction of the cell during short periods, the polarization direction can change, and this creates a random walk for long periods. To model this migration pattern, Langevin-like equations are often used to predict the short-time ballistic motion and the long-time random walk.

On the other hand, a random motion regime has been experimentally observed for very short periods \cite{Selmeczi2005,Dieterich2008,Potdar2009,Gruver2010,Wu2014,Metzner2015,Thomas2020}.
Instantaneous velocities, however, are impossible to be theoretically defined in this regime. The velocity in this case diverges and precludes the use of Langevin dynamics \cite{Rita2022}. 

Additionally, cells do not migrate faster in chemical gradients. Rather, their orientations stabilize in the direction of the gradient, while their drift speeds stay approximately constant. This creates a displacement distribution that does not peak at zero \cite{Masuzzo2017,Nousi2021}. This contradicts the Langevin displacement distribution that peaks exactly at zero.

To address challenges related to diverging velocities, isotropic short-time random walks, and displacement distributions peaked at positive values, we propose a modified single cell migration model with three key components:

First, we introduce the concept of polarization, which guides the cell displacement for periods shorter than the polarization persistent time. Based on experimental observations of correlation between the displacement of the cell and its polarization \cite{Jiang2005,CallanJones2016,Thomas2022}, we define a polarization vector as a geometric object that can be obtained from a single shot image of the cell.

Second, we introduce an isotropic stochastic noise in the cellular position. This comes from the fact that passive  transport lead to actin monomers polymerizing at the front part of the cell \cite{Mogilner2009,Mogilner2020}. This contributes, at least in part, to lamellipodia fluctuations and random walks at short periods. 

Finally, we incorporate a \textit{drift bias} into the model. This preference for forward motion is crucial for emulating the effects of active transport of cytoskeletal components over the cellular front, which results in uniform cellular motion \cite{Mogilner2009,Mogilner2020,Allen2020}. The drift bias is also essential for chemotaxis, as it orients the  polarization of the cell toward a chemical gradient, promoting forward motion. Without the drift bias, the cell would, on average, remain stationary even after aligning its polarization with the chemical gradient.

After establishing the foundational concepts of our topic, we will now provide a clear roadmap for the organization of our paper. We introduce our model for the cell dynamics in the next section. In section 3, we present the main results obtained from this model and compare them with other simulation strategies and experimental results. We conclude this paper in section 4 providing a discussion about the results and some propositions for further exploration.
\section{The Model}
\label{chap:The_Model}
A particle in our model is defined by a position vector $\mathbf{r}$ and a polarization vector $\mathbf{p}(t)=\left(p(t)\cos(\theta),p(t)\sin(\theta)\right)$. In this last equation, $p$ is the polarization intensity and $\theta$ follows a Wiener process\cite{CarloBook} such that:

\begin{eqnarray}
\Delta \theta(t) = \int_{t}^{t+\Delta t} \beta(s)ds \quad \text{,}
\label{eq:theta}
\end{eqnarray}
where $\beta(s)$ is normally distributed with expectation and variance given by:
\begin{eqnarray}
\langle \beta(t) \rangle &=& 0 \nonumber\\
\langle \beta(t)\beta(t^{\prime})\rangle &=& \delta(t-t^{\prime}) \quad \text{.}
\end{eqnarray}

The polarization intensity $p(t)$ after a period $\Delta t$ is written in the form of a Langevin-like equation:
\begin{eqnarray}
   && p(t+\Delta t) = \bigg[(1-\gamma \Delta t) p(t)  \\ &&+ \int_{t}^{t+\Delta t} \left(\xi_{p}(s) + b\right)ds\bigg]
     \hat{p}(t)\cdot\hat{p}(t+\Delta t)\text{,} \nonumber
\label{eq:BAOU_Parallel_Velocity}
\end{eqnarray}
where $\gamma$ corresponds to the cellular loss of polarity, $\xi_{p}$ is a Wiener process with zero expectation and variance given by $g$. Also in the equation, $b$ is a constant bias responsible for inducing a 
forward biased movement of the cell with respect to the polarization axis. The product $\hat{p}(t)\cdot \hat{p}(t+\Delta t)=\cos(\Delta \theta)$, corresponding to a unitary projection of the current polarization onto the polarization axis after a short interval, induces a loss of polarization intensity after the particle reorients.

The displacement of the particle is described by:
\begin{equation}
\Delta \mathbf{r} = \alpha \mathbf{p}(t)\Delta t + \int_{t}^{t+\Delta t}\boldsymbol{\xi}(s)ds \; ,
\label{eq:the_model_particle_displacement}
\end{equation}
where $\alpha$ is a proportionality constant between the cellular polarization and its displacement, ${\boldsymbol{\xi}= \xi_{\parallel} \hat{p} + \xi_{\perp}\hat{n}}$ is a Wiener process that models the phenomenon of fast polymerization and depolymerization of actin in the cytoskeleton. $\hat{p}$ is the orientation of the polarization, while $\hat{n}$ the orientation perpendicular to $\hat{p}$. Finally, ${\langle \xi_{\parallel}(t)\xi_{\parallel}(t^{\prime}) \rangle = \langle \xi_{\perp}(t)\xi_{\perp}(t^{\prime}) \rangle = 2qk\delta(t-t^{\prime})}$ implies that the two components of $\boldsymbol{\xi}$ are uncorrelated. Note that the intensity of $\boldsymbol{\xi}$ is proportional to the intensity of $\beta$, through a proportionality constant $q$. This is due to the fast polymerization-depolymerization phenomenon causing the turnover of both the positional random walk and the cellular polarization.

In summary, our model considers: 1) a point particle with a preferential direction of migration that changes according to the cell polarization reorientation given by the stochastic variable $\beta(t)$. 2) A displacement motion that is parallel to the preferential direction of migration and is proportional to the polarization intensity. This displacement captures the universal coupling between polarization and speed (UCPS)  \cite{CallanJones2016}. 3) An isotropic noise over the cellular displacement generated by $\boldsymbol{\xi}$ whose purpose is to recreate the fast actin polymerization-depolymerization fluctuation.

\begin{figure}[h]
    \centering
    \includegraphics[width=0.35\textwidth]{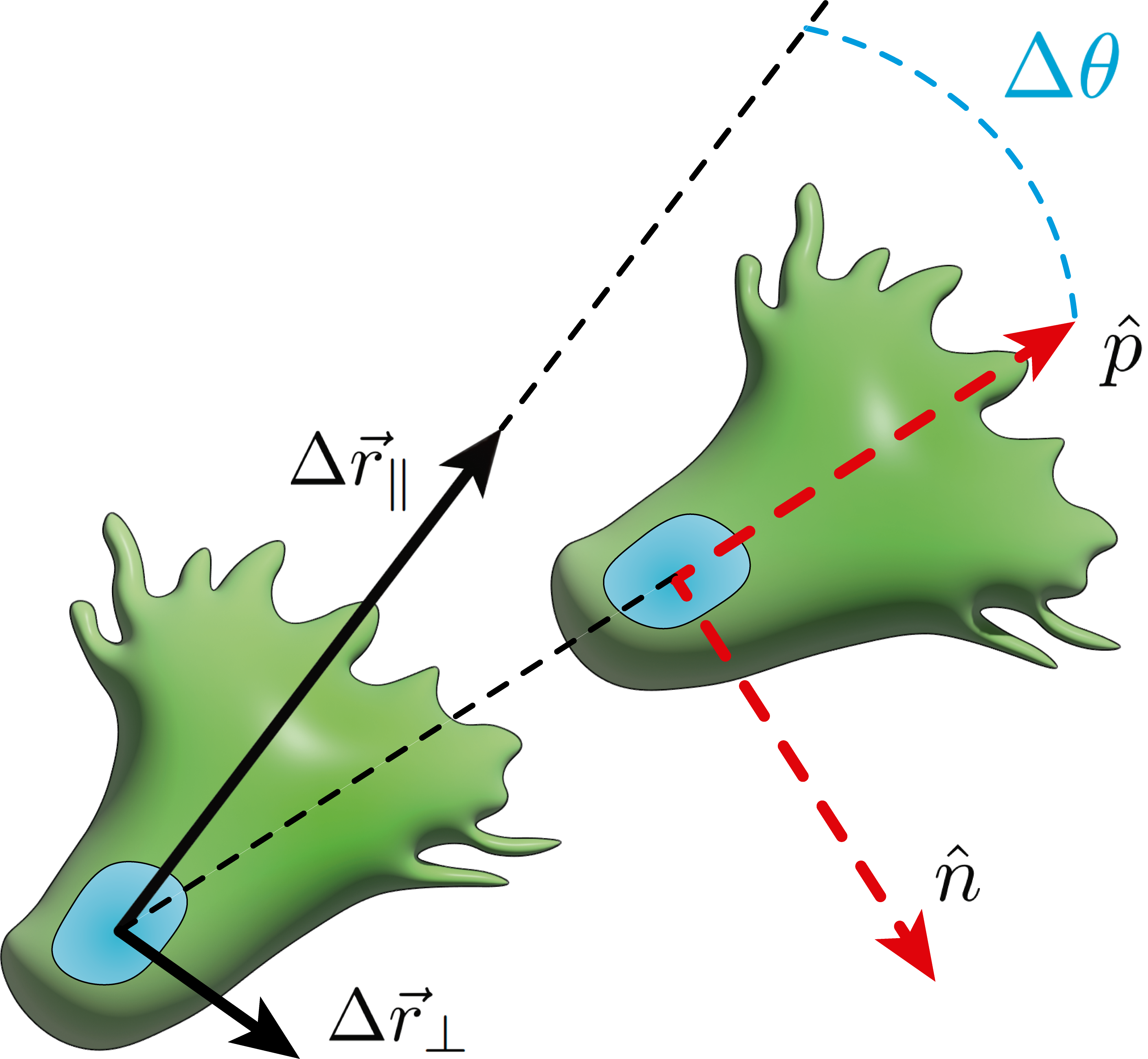}
    \caption{\footnotesize Diagrammatic representation of a particle that moves according to a biased Ornstein-Uhlenbeck process. $\Delta {r}_{\parallel}(t)$ is the displacement of the particle along the polarization axis $\hat{p}=\left(cos(\theta[t]),sen(\theta[t])\right)$ and $\Delta r_{\perp}$ is the displacement along $\hat{n}$, the orientation perpendicular to the polarization.}
 	\label{fig.CellPolarity}
\end{figure}

Figure \ref{fig.CellPolarity} shows a diagrammatic representation of our model.
\section{Analytical Solution}
\label{chap:Analytical_Solution}

The proposed model is solved considering the dynamics of the polarization magnitude for a single time step:
\begin{eqnarray}
\nonumber p(\Delta t) &=& \left((1-\gamma\Delta t) p(0) + \int_{0}^{\Delta t}\left(\xi_{ p}(s) + b\right)ds\right)\\
&& \times  \big(\hat{ p}(0)\cdot \hat{ p}(\Delta t)\big) \quad \text{.}
\label{eq:p_t0}
\end{eqnarray}
This equation is then iterated to time $T = n\Delta t$ with $n \in \mathbb{N}$:
\begin{eqnarray}
\nonumber && p(n\Delta t) = (1-\gamma\Delta t)^{n} p(0)\prod_{i=0}^{n-1} (\hat{ p}_{i}\cdot\hat{ p}_{i+1}) \\
\nonumber && + \sum_{i=1}^{n} (1-\gamma \Delta t)^{n-i}\left(\int_{(i-1)\Delta t}^{i\Delta t}(\xi_{ p}(s)+b)ds\right)\\
&& \times \prod_{j=i-1}^{n-1}(\hat{ p}_{j}\cdot\hat{ p}_{j+1}) \quad \text{,}
\label{eq.par_pol_N}
\end{eqnarray}
where the sub-index notation ${\hat{p}(n\Delta t) = \hat{ p}_{n}}$ is used for simplicity.
Every integral with respect to Wiener processes are  Ito-integrable, which, can be represented as iterations. 

The expected value of equation (\ref{eq.par_pol_N}) is:
\begin{eqnarray}
    \langle  p(T)\rangle=\frac{b}{\gamma+k} + \left( p(0) - \frac{b}{\gamma +k}\right) e^{-(\gamma + k)T} \quad \text{,}
\label{eq.avg_par_vel_lim}
\end{eqnarray}
whose stationary average polarization is $p_{\text{stationary}}=\frac{b}{\gamma+k}$, corresponding to the polarization in the limit where ${T \to \infty}$. The variance of the polarization is obtained from its standard definition ${\langle p^{2}(N\Delta t) \rangle - \langle p(N\Delta t) \rangle^{2}}$, giving:
\begin{eqnarray}
    \langle p^{2}(T)\rangle - \langle  p(T)\rangle^{2}=\frac{g}{2(\gamma+k)}\left(1-e^{-2(\gamma+k)T}\right) \text{.}
    \label{eq.pol_variance}
\end{eqnarray}

Equations (\ref{eq.avg_par_vel_lim}) and (\ref{eq.pol_variance}) evidence an asymptotic polarization distribution with constant variance $\frac{g}{2(\gamma+k)}$ and positive mean. Stated differently, the particle has a bias towards forward movement, since polarization and displacement are correlated. We call this phenomenon, the \textit{drift bias}.

The particle's position $\mathbf{r}(T)$ is the result of combining equations \ref{eq:the_model_particle_displacement} and \ref{eq.par_pol_N} for an arbitrary time-step $n\Delta t$ and then summing it from $t=0$ to $T=N \Delta t$. This results in:
\begin{eqnarray}
    \nonumber && \mathbf{r}\left(T\right) =\mathbf{r}(0) \\
    \nonumber &&+ \alpha  \Delta t\;  p(0) \sum_{n=0}^{N-1}(1-\gamma \Delta t)^n \prod_{i=0}^{n-1} (\hat{ p}_{i} \cdot \hat{ p}_{i+1}) \hat{ p}_{n} \\
    \nonumber &&+ b\alpha \Delta t^2 \sum_{n=0}^{N-1}\sum_{j=0}^{n-1}(1-\gamma \Delta t)^{n-1-j}\prod_{i=j}^{n-1}(\hat{ p}_{i} \cdot \hat{ p}_{i+1}) \hat{ p}_{n} \\
    \nonumber &&+ \alpha\Delta t \; \sum_{n=0}^{N-1} \sum_{j=0}^{n-1} (1-\gamma \Delta t)^{n-1-j} \prod_{i=j}^{n-1}(\hat{ p}_{i} \cdot
    \hat{ p}_{i+1})\\
    \nonumber && \times \int_{j \Delta t}^{(j+1)\Delta t} \xi_{ p}(s) ds \,\hat{ p}_{n}\\
    \nonumber &&+ \alpha\Delta t \; \sum_{n=0}^{N-1}\int_{n\Delta t}^{(n+1)\Delta t}  [(n+1)\Delta t-s]\xi_{ p}(s) ds\, \hat{ p}_{n} \\
    &&+ \sum_{n=0}^{N-1} \int_{n\Delta t}^{(n+1)\Delta t} \boldsymbol{\xi}(s)\, ds \quad \text{.}
    \label{eq.cap3_Particle's_General_Position}
\end{eqnarray}

Equations (\ref{eq.par_pol_N}) and (\ref{eq.cap3_Particle's_General_Position}) fully characterize the dynamics of our model through important functions such as the mean square displacement (\textit{MSD}), the polarization autocorrelation function (\textit{PACF}), the mean velocity auto-correlation function (\textit{mVACF}) and the average velocity distribution function (\textit{VPDF}). All of which will be explained next. 

The polarization autocorrelation function, defined as ${ PACF(\Delta T) = \langle \vec{p}(T)\cdot\vec{p}(T+\Delta T) \rangle }$, where the expectation is calculated over different noise realizations for every period $T$:
\begin{eqnarray}
    \nonumber PACF (\Delta T) &=& \frac{g}{2(\gamma+k)}e^{-(\gamma+2k)\Delta T}\\
                              &+& \frac{b^{2}}{(\gamma+k)^{2}}e^{-k\Delta T}\quad \text{.}
    \label{eq:PACF}
\end{eqnarray}
In short, the polarization of the particle has a memory and it decays as the sum of two exponential functions. 

Twice differentiating Eq. (\ref{eq:PACF}) with respect to $\Delta T$ yields the mean square displacement (\textit{MSD}), a measure that distinguishes ballistic from random walk regimes. Its analytical solution is written as:
\begin{eqnarray}
\nonumber MSD &=& 2D_{1}\left[ \frac{\Delta T}{1-S} - P_{1}\left(1-e^{-\Delta T/P_{1}}\right)\right]\\
&+& 2D_{2}\left[\Delta T - P_{2}\left(1-e^{-\Delta T/P_{2}}\right)\right]\text{,}
\label{eq:Biased_MSD_Solution}
\end{eqnarray}
where $P_{1,2}$ are persistence coefficients, $D_{1,2}$ are diffusion coefficients, and $S$ represents the period where short-time random walk occurs. We call it the \textit{excess diffusion coefficient}. Each coefficient is written in terms of the parameters of the model as:
\begin{eqnarray}
    D_{1} &=& \frac{g\alpha^{2}}{2(\gamma+k)(\gamma+2k)} \label{eq:D1}\\
    P_{1} &=& \frac{1}{(\gamma+2k)} \label{eq:P1}\\
    D_{2} &=& \frac{b^{2}\alpha^{2}}{k(\gamma+k)^{2}} \label{eq:D2}\\
    P_{2} &=& \frac{1}{k} \label{eq:P2}\\
    S &=& \frac{4qk(\gamma+k)(\gamma+2k)}{g\alpha^{2} + 4qk(\gamma+k)(\gamma+2k)}\quad \text{.}
    \label{eq:excess_diffusion}
\end{eqnarray}

The relation between the coefficients of diffusion, persistence, \textit{excess diffusion} and the parameters of our model $\gamma,g,k,b,q$ can be defined in many ways. However, relations \ref{eq:D1}-\ref{eq:excess_diffusion} were written with the purpose of maintaining consistency with previous models \cite{Rita2022}. Also, they facilitate the interpretation of each of the five coefficients, which can be determined experimentally. The experimental procedures for characterization in terms of the parameters of our model are explained at the end of this chapter.

We rewrite the \textit{MSD} in its natural units, with ${\langle|\boldsymbol{\rho}|^{2}\rangle = MSD/2D_{1}P_{1} = \langle|\mathbf{r}|^{2}\rangle/2D_{1}P_{1} }$. This gives:
\begin{eqnarray}
    \nonumber \langle |\boldsymbol{\rho}|^{2} \rangle &=& \frac{\tau}{1-S} - (1- e^{-\tau})\\
    &+& \lambda\phi\left[\frac{\tau}{\phi} - (1-e^{-\tau/\phi})\right] \quad \text{.}
    \label{eq:Natural_Units_MSD}
\end{eqnarray}

Expression (\ref{eq:Natural_Units_MSD}) contains three parameters in the natural unit system: the persistence ratio ${ \phi=P_{2}/P_{1} }$, the diffusion ratio ${ \lambda=D_{2}/D_{1} }$, and the excess diffusion coefficient $S$.

The analytical \textit{MSD} solution not only agrees with the numerical results of our model but also with CompuCell3D migration modelling results\cite{Fortuna2020}. Where CompuCell3D is a Potts Modelling software \cite{Graner1992} capable of reproducing the dynamics of certain systems through Monte Carlo energy minimization algorithms, a completely different approach from ours, but that has produced similar results. The model also supports the existence of the three observed migration regimes.  They are characterized, in natural units, as random walk for $\tau < S$, predominant ballistic motion for $S < \tau < max(1,\phi)$, and another random walk for $\tau > max(1,\phi)$. The persistence time changes according to $\phi$. If $\phi<1$, then $e^{-\Delta\tau}$ is more significant than $e^{-\Delta\tau/\phi}$. However, if $\phi>1$, then $e^{-\Delta\tau/\phi}$ becomes the more important term, extending the time of persistence. These patterns have also been observed in various experiments conducted by different laboratories \cite{Selmeczi2005,Dieterich2008,Potdar2009,Gruver2010,Wu2014,Metzner2015,Thomas2020}. 

\begin{figure}[H]
    \includegraphics[width=0.37\textwidth, angle=270]{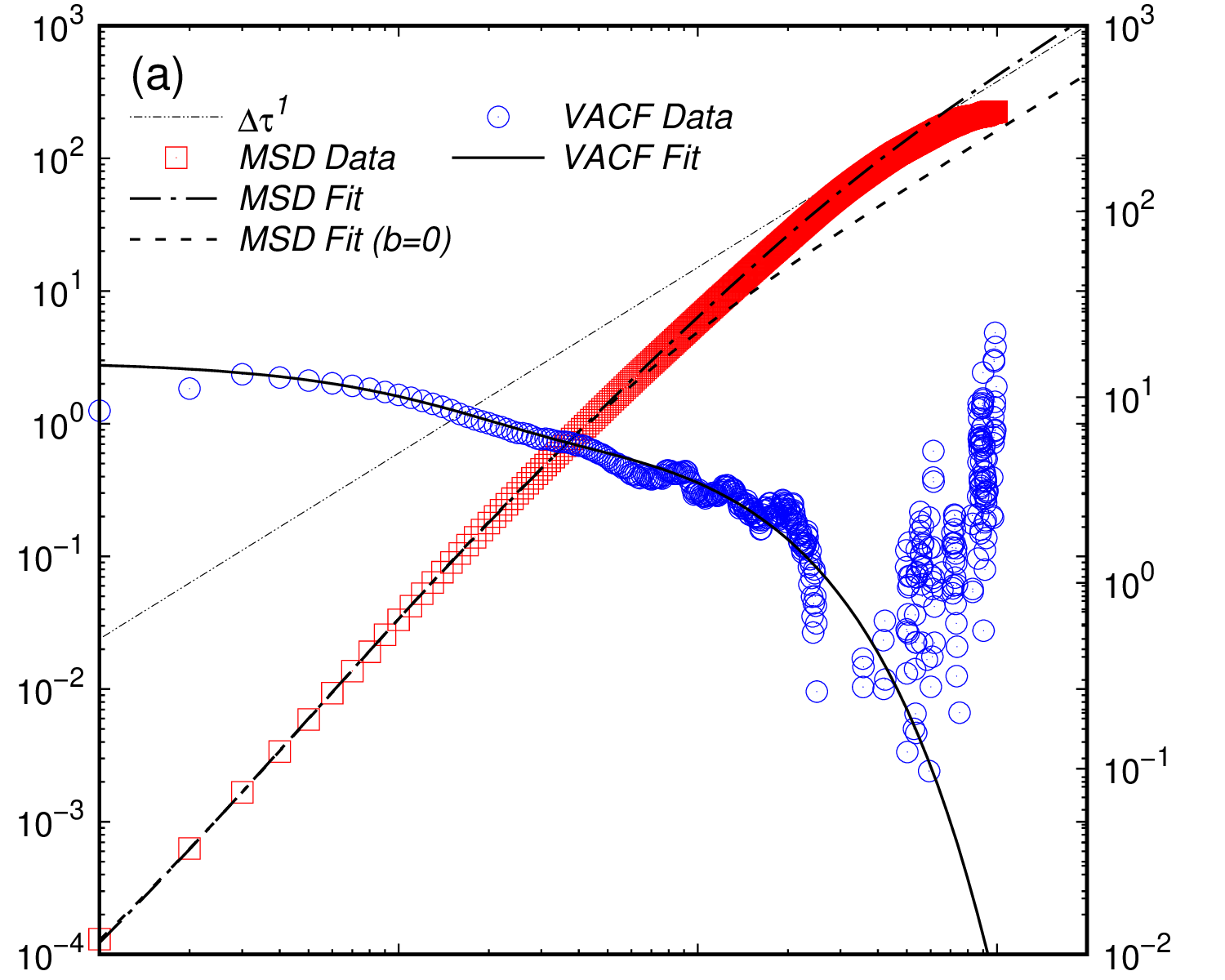}
    \centering
    \includegraphics[width=0.345\textwidth, angle=270]{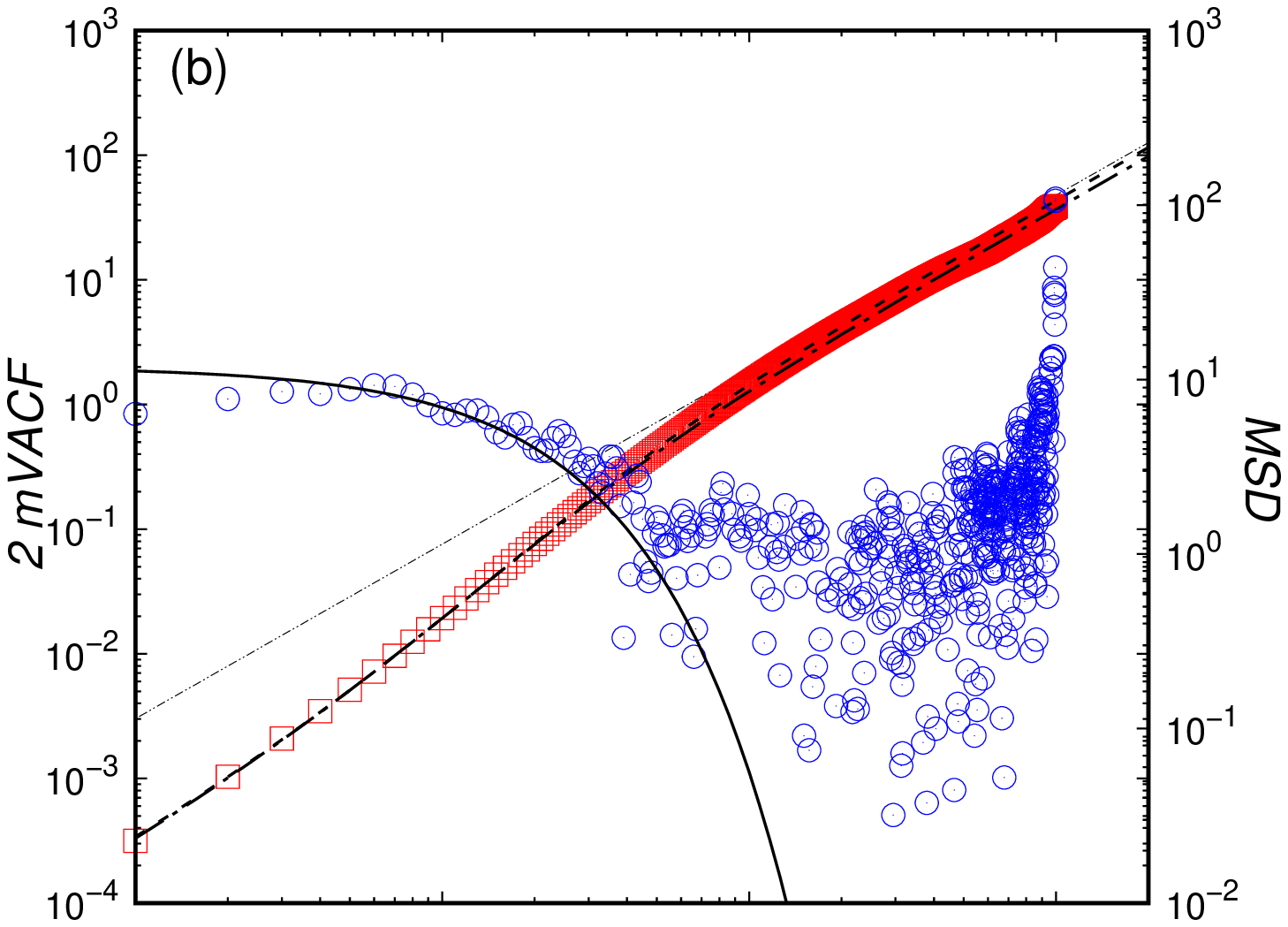}
    \includegraphics[width=0.37\textwidth, angle=270]{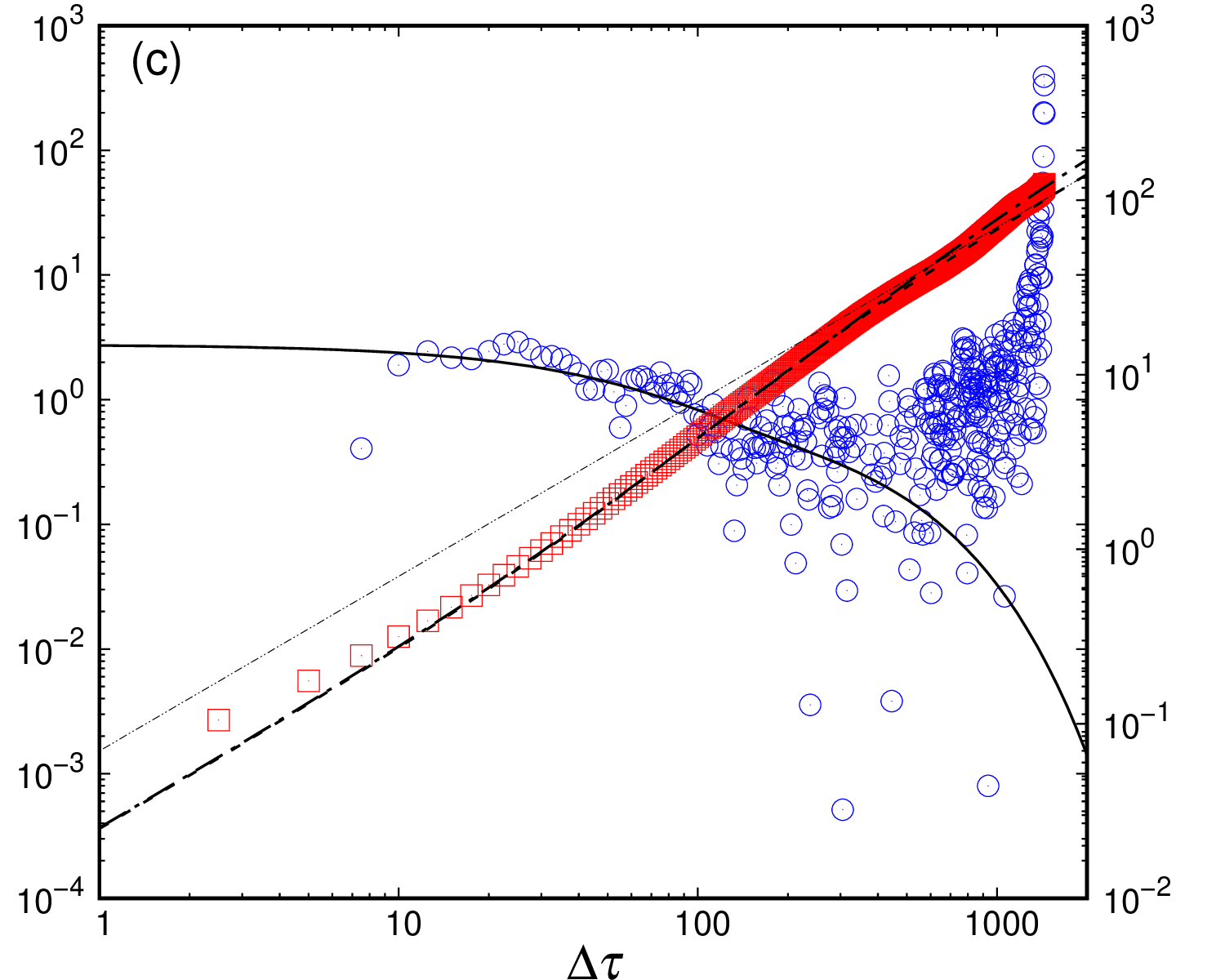}
    \caption{\footnotesize Comparison of experimental migration data and our single cell migration model in a log-log scale, where data was produced by \cite{Metzner2015}.The trajectories are obtained from the breast carcinoma cell line MDA-MB-231 for three different substrate materials: (a) Fibronectin coated plastic, (b) uncoated plastic and (c) collagen coated plastic. The legend indicates the following: The time dependence of experimental \textit{MSD} and \textit{mVACF} data in natural units are plotted on log-log axes.}
    \label{fig:Experimental_Data_MSD_mVACF}
\end{figure}

\begin{figure}
    \ContinuedFloat
    \caption*{\footnotesize In the figure, squares represent \textit{MSD} data and dots represent \textit{mVACF} data. The first dotted line with $\Delta \tau^{1}$ indicates a pure random walk with unitary slope. The line labeled \textit{MSD Fit (b=0)} corresponds to a fit assuming zero drift bias ($D_{2}=0$). The dotted line labeled \textit{MSD Fit} represents the complete biased \textit{MSD} fit. The continuous line represents the \textit{mVACF} fit.}
\end{figure}

Using a normalized root mean square error (\textit{nRMSE}) to quantify the quality of the fits, written as
\begin{equation}
    nRMSE = \frac{100\%}{max(\mathbf{\hat{y}})-min(\mathbf{\hat{y}})}\sqrt{\sum_{i=1}^{N}\frac{(y_{i}-\hat{y}_{i})^{2}}{N}}
\end{equation}
where $\mathbf{\hat{y}}$ represents the experimental data-set, $\hat{y}_{i}$ represents each of component of the experimental data-set and $y_{i}$, the predicted values using our model and N is the size of the experimental data-set. With this, we are able to quantify the quality of the fits in percent terms with respect to the range of experimental values.

We obtained the following:
\begin{table}[h]
    \centering
    \begin{tabular}{| c | c | c |}
        \hline
        Substrate Type & \textit{MSD} & \textit{mVACF} \\
        \hline
        Fibronectin & 0.68819  \% & 0.08076 \%\\
        \hline
        Collagen & 5.35556  \% & 0.008507 \%\\
        \hline
        Plastic & 0.05295 \% & 0.18487 \%\\
        \hline
    \end{tabular}
    \caption{Biased fit \textit{nRMSE}}
    \label{tab:Quality_of_Fit_Biased}
\end{table}
and for the unbiased fits:
\begin{table}[h]
    \centering
    \begin{tabular}{| c | c | c |}
        \hline
        Substrate Type & \textit{MSD} & \textit{mVACF}\\
        \hline
        Fibronectin & 0.78252 \% & 0.19311 \%\\
        \hline
        Collagen & 5.46699 \% & 0.00807 \%\\
        \hline
        Plastic & 0.20151 \% & 0.18794 \%\\
        \hline
    \end{tabular}
    \caption{Unbiased fit \textit{nRMSE}}
    \label{tab:Quality_of_Fit_Unbiased}
\end{table}

Figure \ref{fig:Experimental_Data_MSD_mVACF} illustrates the steps for fitting experimental data for cell migration using the Levenberg-Marquardt algorithm \cite{Levenberg1944,Marquadt1963} in Gnuplot 5.4. The fitting process involves the following steps: (i) Estimating the $D_{2}$ parameter by analyzing the \textit{mVACF} value at the beginning of its second regime. (ii) Exponential fitting the decay ratio $P_{2}$ and the initially estimated parameter $D_{2}$ for the second \textit{mVACF} regime.  (iii) Fitting the complete bi-exponential function (\ref{eq:Mean_Velocity_Auto-Correlation_Solution_Natural_Units}) to determine $D_{1}$, $P_{1}$, $D_{2}$, and $P_{2}$. (iv) Fitting the \textit{excess-diffusion coefficient} $S$ by fitting the \textit{MSD} function (\ref{eq:Biased_MSD_Solution}). 

Our model fits the calculated \textit{MSD} and \textit{mVACF} with experimental data better than our previous model based on an Ornstein-Uhlenbeck process \cite{Rita2022} see tables (\ref{tab:Quality_of_Fit_Biased},\ref{tab:Quality_of_Fit_Unbiased}) for the error comparison, see that for the fibronectin substrate we verify a reduction of more than half of the \textit{nRMSE}. Moreover, the plots shown in Fig. \ref{fig:Experimental_Data_MSD_mVACF} indicate that the experimental \textit{mVACF} exhibit two different decay rates. Our model successfully captures this behavior. This is because the timings corresponding to a change in the polarization direction and the cell movement via acto-myosin mechanisms differ significantly, but equally affect the \textit{mVACF}. Whereas the change in polarization affects the \textit{mVACF} via memory loss, the cell movement affects the \textit{mVACF} through short-time fluctuations in the cellular position.

There are two points, however, that should be clarified, the first is that Fig. (\ref{fig:Experimental_Data_MSD_mVACF}) plot (c) shows an experimental \textit{mVACF} with a slight decay for $\Delta \tau \to 0$, which indicates a sub-diffusion process and may be indicative of anomalous migration \cite{Selmeczi2005, Metzner2015} something that is not encompassed by our model but that could be incorporated if desired. The second point is that in Fig. (\ref{fig:Experimental_Data_MSD_mVACF}) plot (b), there where two stable fits that we managed to obtain, one that is shown in Fig. (\ref{fig:Experimental_Data_MSD_mVACF}) and another that would produce a slightly better quality fit for the \textit{mVACF} but worse values for the \textit{MSD} (\textit{nRMSE}=0.18794\%) note however that in terms of the percent values of the total data range, all fits present very low root mean square errors.

The characterization of cellular dynamics is nearly complete by first measuring cell trajectories and time-series for the polarization, and then evaluating the \textit{MSD} and \textit{PACF}. However, there are not currently standardized protocols for measuring polarization time-series\cite{Thomas2022}. Therefore, the following discussion explains how to extract the polarization information from the trajectories. 

Instantaneous displacements cannot be experimentally measured, since each successive microscope photography takes a finite period. Thus, although the position and polarization should ideally define the cellular dynamics fully, this interval should be accounted for. This issue is addressed by defining an average velocity as the ratio of a particle's displacement in Eq. \ref{eq:the_model_particle_displacement} and the elapsed time $\delta$ between consecutive measurements:
\begin{eqnarray}
    \vec{v}_{\text{avg}}(\delta,t) = \frac{\Delta \vec{r}}{\delta} = \alpha\frac{\vec{p}(t)}{\delta}\Delta t + \int_{t}^{t+\Delta t} \frac{\vec{\xi}(s)}{\delta}ds.
\end{eqnarray}
Because $\vec{\xi}$ follows a Wiener process, the average velocity will exhibit different regimes described as follows: For very small values of $\delta<SP_{1}$, the ratio ${|\vec{\xi}|/\delta}$ is significantly bigger than ${|\vec{p}|/\delta}$, resulting in a predominantly random walk behavior and an average velocity  ${\vec{v}_{\text{avg}} \approx \int{t}^{t+\Delta t} \Delta\vec{r}/\delta ds}$. When $\delta$ is sufficiently large ($SP_{1}<\delta<\max(P_{1},P_{2})$), ${ |\vec{p}|/\delta >> |\vec{\xi}| /\delta }$. Consequently, the approximate average velocity becomes ${ \vec{v}_{\text{avg}}\approx \alpha\vec{p}\Delta t /\delta }$, which converges to a finite value. Such observation leads to a reliable definition of average velocity when a particle exhibits short-time random walk. We define the latter as the ratio of displacement and time during the interval when migration is mostly ballistic. This interval corresponds to the steepest slope in the mean square displacement (\textit{MSD}) function.

Additionally, it is worth noting that for an intermediate value of $\delta$, the auto-correlation function of $\vec{v}_{\text{avg}}$ is isomorphic to the polarization auto-correlation function (PACF), but scaled by the proportionality constant $\alpha^2$. This leads us to the concept of the mean velocity auto-correlation function (\textit{mVACF}), denoted as $\Psi$, which factors in the finite time interval $\delta$ for each displacement value $\Delta \vec{r}$.

The analytical solution is found by combining equations (\ref{eq.cap3_Particle's_General_Position}) and:

\begin{equation}
    \psi(\varepsilon,\Delta \tau) \equiv \langle \vec{u}_{\text{avg}}(\tau,\varepsilon) \cdot \vec{u}_{\text{avg}}(\tau+\Delta \tau,\varepsilon) \rangle,
\end{equation}
in natural units, where ${\varepsilon\equiv \delta/P_{1}}$ and ${ \vec{u}_{\text{avg}}\equiv \vec{v}_{\text{avg}}\sqrt{P_{1}/D_{1}} }$. The solution is given by:
\begin{eqnarray}
    \nonumber \psi_{\varepsilon,\Delta \tau} &=& e^{-\Delta \tau} \frac{\left(\cosh\big(\varepsilon\big) - 1\right)}{\varepsilon^{2}}\\
    &+& \frac{\lambda}{\phi} e^{-\Delta \tau/\phi} \frac{\left(\cosh\big(\varepsilon/\phi\big) - 1\right)}{(\varepsilon/\phi)^{2}},
    \label{eq:Mean_Velocity_Auto-Correlation_Solution_Natural_Units}
\end{eqnarray}
where the fractions after the exponential terms, with an hyperbolic cosine, can be interpreted as a correction factor that goes to $1$ as $\varepsilon \to 0$ and $\psi(\varepsilon,\Delta \tau) \overset{\varepsilon\to 0}{=}\langle \vec{\Pi}(\tau+\Delta\tau)\cdot\vec{\Pi}(\tau) \rangle$, where $\vec{\Pi}$ is the polarization vector in natural units. The \textit{mVACF} is proportional to \textit{PACF} when the time between each successive displacement is exactly zero. The correction factor allows comparison between experimental displacement auto-correlation measurements with polarization auto-correlation obtained with our model even without having a polarization time-series data available.

The \textit{MSD}, \textit{PACF} and \textit{mVACF} measurements distinguish random walk from ballistic regimes. However, these values are unable to characterize the \textit{drift bias} or directional anisotropies such as chemoattraction. To model this effect, we also calculated the probability distribution function (\textit{PDF}) for cellular displacements. Our earlier model, as presented in \cite{Rita2022}, featured a parallel-to-polarization probability density function (\textit{PDF}) that peaks at zero. However, this outcome contradicts both single-cell migration experiments and the findings in chemotaxis, where particles are required to align themselves with a chemical gradient and exhibit forward movement as their preference. Therefore, it is necessary to include a bias in the  Langevin equation corresponding to the polarization. 

An analytical \textit{PDF} for the particle's polarization magnitude is obtained as follows. Assuming an infinitesimal variation in the \textit{PDF} ${\Delta \varrho\big( \Pi(\tau)\big) = \varrho\big( \Pi(\tau+\Delta \tau)\big) - \varrho\big( \Pi(\tau)\big)}$, we expand $\varrho\big( \Pi(\tau+\Delta \tau)\big)$ in a Taylor series and consider only the terms of second order or less in $\Delta t$. With this, we obtain the Fokker-Planck equation\cite{CarloBook} in natural units: 
\begin{eqnarray}
    \nonumber \frac{\partial  \varrho(\Pi,\tau)}{\partial \tau} &=& \frac{\phi-1}{\phi}\Bigg[\frac{\partial }{\partial \Pi}\Big( \varrho(\Pi,\tau)(\Pi - 1)\Big) \\ 
    &+& \frac{\phi}{\lambda}\frac{\partial^{2} \varrho(\Pi,\tau)}{\partial \Pi^{2}} \Bigg] \quad \text{,} \label{eq:Fokker_Planck}
\end{eqnarray}
where $\Pi = \alpha |\vec{p}| \sqrt{P_{2}/D_{2}}$ is the polarization magnitude in natural units, $\phi$ and $\lambda$ are, respectively, the persistence and diffusion ratios. The stationary solution ($\frac{\partial \varrho(p,t)}{\partial t} = 0$) for the Fokker-Planck equation (\ref{eq:Fokker_Planck}) is a Gaussian function centered at $\Pi = 1$ or $p = b/(\gamma+k)$. This means that the particle has the highest probability of displacement  proportional to ${ \Delta r_{\parallel}=\alpha \frac{b}{\gamma+k}\Delta t }$, similar what is observed  experimentally. 

The polarization PDF can be experimentally obtained by measuring the cellular displacements during intervals when the ballistic regimes are mostly pronounced. The numerical results are shown in Fig. (\ref{fig:avg_v_pdf}).

\newpage 
\begin{strip}
% \begin{center}
\begin{minipage}{\textwidth}
\begin{figure}[H]
    \centering
    \includegraphics[width=\textwidth]{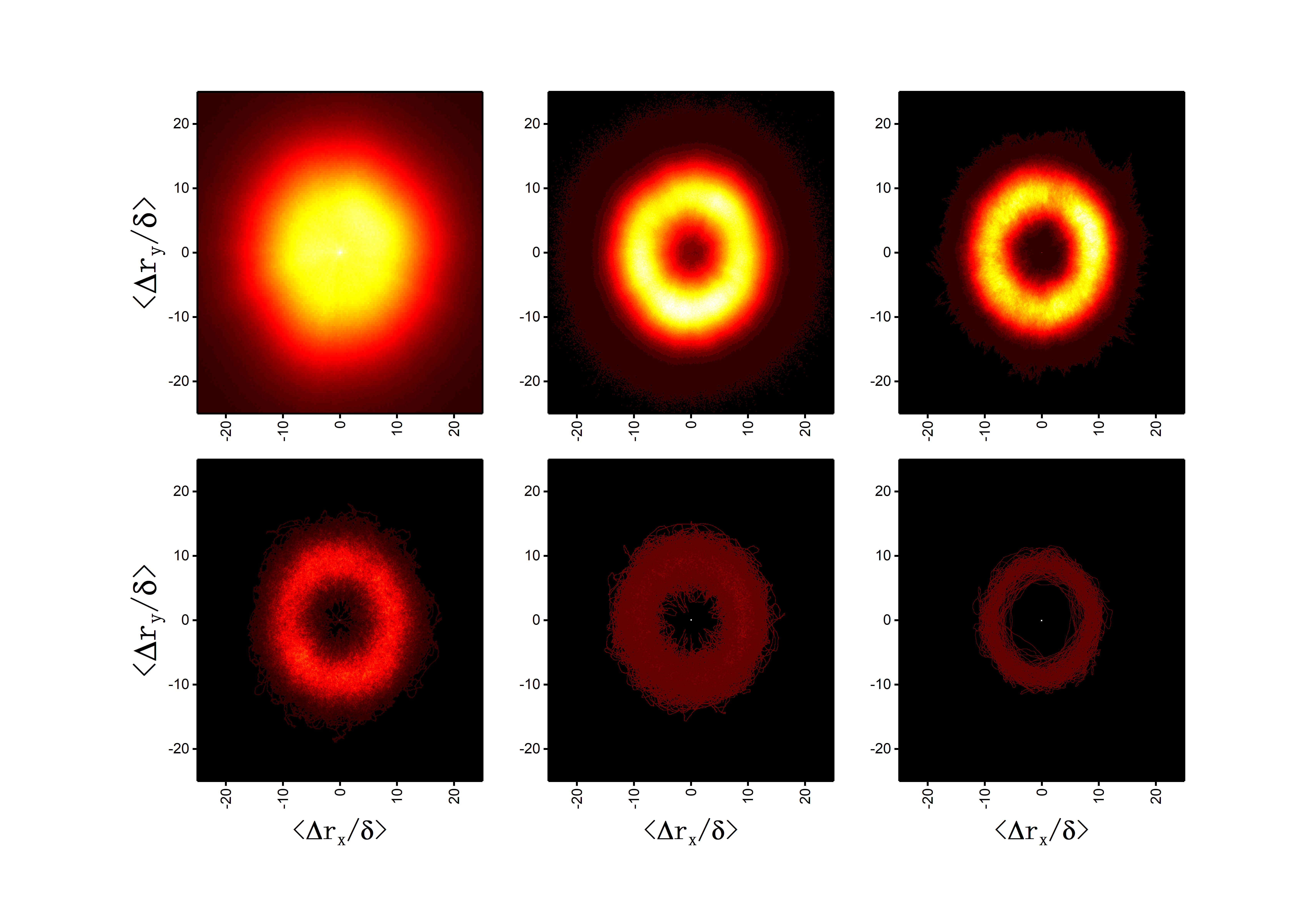}
    \caption{\footnotesize Top view of the mean velocity probability density function obtained from the numerical solutions, for  varying values of $\delta$ ($\delta = 0.0001, \, 0.001, \, 0.01, \, 0.1, \, 1, \, \text{and} \, 10 $ from left to right and top to bottom). The data considers $10$ particles and parameter values $g=10$, $q=0.1$, $\gamma=1$, $k=0.04405$, and $b=10$. Brightness corresponds to the number of occurrences. Each  parameter was chosen to maintain consistency with our previous model \cite{Rita2022}.}
    \label{fig:avg_v_pdf} 
\end{figure}
\end{minipage}
% \end{center}
\end{strip}

Figure \ref{fig:avg_v_pdf} demonstrates that adjusting the time interval $\delta$ in the calculation $\vec{v}_{\text{avg}} = \frac{\Delta \vec{r}}{\delta}$ affects both the shape and magnitude of the average velocity distributions. The first figure depicts a distribution with a peak at zero and a large variance, indicating that velocity values diverge when $\delta$ corresponds to a random walk regime. The second, third, and fourth figures show distributions with hollow centers and smaller variances, suggesting average velocity values centered around $\alpha b/(\gamma+k)$. The average velocity is obtained from solving the stationary Fokker-Planck equation considering all possible orientations. The fifth and sixth figures show distributions where most points lie at the center due to a significantly larger time interval as compared to the  ballistic regime. This results in average velocities close to zero.

The results above show that the time interval has a high impact in the quality of the collected average velocity data and that there is an optimal time interval for measuring  average velocities. An ideal experimental procedure for precise velocity measurements is to film the trajectories of some cells, obtain the mean square displacement and, through the analysis of the exponents of the mean square displacement curve, find the ideal time interval when the trajectory is mostly ballistic. The average displacement distribution should converge to the solution of equation \ref{eq:Fokker_Planck}, a non-zero Gaussian distribution. 

The single cell migration model has five parameters, each modeling a certain cellular behavior. To compare the model with experiments, five different experimental values are needed. They may be determined by measuring the \textit{MSD}, \textit{mVACF} and the average velocity \textit{PDF}. Even though we added one more parameter to describe the system, namely the drift bias, it does not add complications to the experimental analysis. The steps needed to compare the model with experiments are:

1. Acquire complete trajectory data and compute the resulting \textit{MSD} to distinguish between distinct migration regimes: fast-time random walk, followed by the ballistic intermediate-time regime, and ultimately, long-time random walk.

2. Analyze the \textit{MSD} curve to identify the time interval associated with the most pronounced ballistic migration regime, which is indicated by the steepest slope in the \textit{MSD}. This process enables the determination of the optimal measurement interval for defining average velocities. 

3. Utilize the optimal time interval to calculate cellular displacements and generate a distribution of average velocities. Subsequently, evaluate the mean and variance of the distribution, which in the model are represented by ${ B=\frac{b^{2}\alpha^{2}}{(\gamma+k)^{2}} }$ and ${ G = \frac{g\alpha^{2}}{2(\gamma+k)} }$, respectively.

4. Compute the values of the \textit{mVACF} and observe its rate of decay. If only one rate of exponential decay is observed, it becomes challenging to distinguish the noise component from the drift bias. In such case, the experimental analysis should follow the approach described in \cite{Rita2022}. However, if two distinct regimes can be identified, as in Fig. (\ref{fig:Experimental_Data_MSD_mVACF}), fit the \textit{mVACF} solution to data to extract the two decay rates and the associated coefficients $D_{1},P_{1},D_{2},P_{2}$.

5. Fit the \textit{MSD} function \ref{eq:Biased_MSD_Solution} to the experimental data together with the previously obtained parameters to determine the \textit{excess diffusion coefficient} $S$.

6. Determine the ratio between the persistence coefficients $\phi = \frac{P_{2}}{P_{1}}$ and the diffusion coefficients $\lambda = \frac{D_{2}}{D_{1}}$ and calculate the \textit{MSD} in natural units.

Such procedure fully characterizes experimental data and allows one to compare simulations of different cell strands and single cell dynamics. The only unexplained phenomenon in our model is a correlation decay in \textit{mVACF} for small $\Delta T$ values. In \cite{Rita2022}, we explained that taking the average of a squared cell displacement that is consistent with a well defined converging part and a random walk part, would incur in loss of correlation. While this theory, due to finite precision measurements, should manifest experimentally, it is intriguing that the decrease in correlation is also observed in CompuCell3D cell simulations, which have sufficient precision. This similarity in correlation decay for $\Delta T\to 0$ suggests that, while part of the decay may be attributed to the finite precision of measurements, there must exist another contributing factor \cite{Selmeczi2005}.

\section{Discussions and Concluding Remarks}
\label{chap:Conclusions}

We have shown that a model that incorporates spatial anisotropy can accurately simulate cellular motion. Cells, however, have asymmetric cytoskeletal structures that facilitate their migration, which makes orientation anisotropy alone fail to replicate experimental results. Each cell also experience  noise in all directions, which arises from the passive transport of actin filaments within its interior\cite{Mogilner2009,Mogilner2020}. Consequently, defining instantaneous velocities becomes challenging. To analyze cellular motion effectively, we propose a quantity that remains well-defined, regardless of the time interval used to measure cellular displacements.

Previous research \cite{CallanJones2016} shows that the correlation between cellular displacements and cell polarization is well established. Cell polarization, characterized by a geometrical quantity given by a vector connecting the cellular center of mass to its nucleus, always remains well-defined. By considering this polarization for measuring displacement, we were able to introduce noise into the particle's movement. This approach maintains a well-defined displacement while promoting a biased forward motion, which enables migration under a chemical field. Remarkably, this model reproduces the three migration regimes observed experimentally.

In cases where the polarization data is not available, one can still analyze cellular dynamics satisfactorily by determining the optimal time interval (corresponding to the highest slope in the \textit{MSD}) that ensures a well-defined and converging average velocity. Subsequently, calculations of the \textit{mVACF} and the average velocity \textit{PDF} can provide more insights. These measurements quantify crucial aspects of cellular dynamics, such as the extent of position fluctuations, the typical average velocity of cell migration, the way polarization orientation changes, and the transition from a random walk to a ballistic behavior in a trajectory.

In conclusion, our model captures most of the characteristics of single cell dynamics. And although there are some different approaches such as active bio-fluid models (see for instance \cite{Martin2021,Fodor2021}), these techniques would still not be sufficient to reproduce specific individual behaviors described by our model, even if the case of non interacting fluid particles. For instance, our model provides a mean velocity distribution where the peak assumes positive values. Moreover, the mean velocity autocorrelation function (mVACF) in our model shows a bi-exponential decay, implying processes that decrease at different rates. This behavior depends on timing of processes related to the polarization orientation change and the internal cytoskeletal rearrangement which exhibit different characteristic time-scales.

\nocite{*} %Reference without citing
\bibliographystyle{naturemag}
\bibliography{bib}

\end{document}